\documentclass[twocolumn]{aastex63} 

\hypersetup{linkcolor=red,citecolor=blue,filecolor=cyan,urlcolor=blue}

\usepackage{graphicx,color}
\usepackage{amssymb}
\usepackage{amsmath}
\usepackage{url}
\usepackage{natbib}
\usepackage{txfonts}

\newcommand{\FeXVIII}{\ion{Fe}{18}}

\usepackage{url}
\usepackage{hyperref}

\begin{document}
	
\title{Dominance of Bursty over Steady Heating of the 4--8 MK Coronal Plasma in a Solar Active Region: Quantification using Maps of Minimum, Maximum, and Average Brightness}
	
	\author[0000-0001-7817-2978]{Sanjiv K. Tiwari}
	\affiliation{Lockheed Martin Solar and Astrophysics Laboratory, 3251 Hanover Street, Bldg. 252, Palo Alto, CA 94304, USA}
	\affiliation{Bay Area Environmental Research Institute, NASA Research Park, Moffett Field, CA 94035, USA}
	
	\author{Lucy A. Wilkerson}
	\affiliation{University of Maryland, College Park, MD 20742, USA}

	\author[0000-0001-7620-362X]{Navdeep K. Panesar}
	\affiliation{Lockheed Martin Solar and Astrophysics Laboratory, 3251 Hanover Street, Bldg. 252, Palo Alto, CA 94304, USA}
	\affiliation{Bay Area Environmental Research Institute, NASA Research Park, Moffett Field, CA 94035, USA}

    \author[0000-0002-5691-6152]{Ronald L. Moore}
    \affiliation{NASA Marshall Space Flight Center, Mail Code ST 13, Huntsville, AL 35812, USA}
    \affil{Center for Space and Aeronomic Research, The University of Alabama in Huntsville, Huntsville, AL 35805, USA}
    
    \author[0000-0002-5608-531X]{Amy R. Winebarger} 
    \affiliation{NASA Marshall Space Flight Center, Mail Code ST 13, Huntsville, AL 35812, USA}

	\begin{abstract}
A challenge in characterizing active region (AR) coronal heating is in separating transient (bursty) loop heating from the diffuse background (steady) heating.  We present a method of quantifying coronal heating's bursty and steady components in ARs, applying it to \FeXVIII\ (hot94) emission of an AR observed by SDO/AIA.  The maximum, minimum, and average brightness values for each pixel, over a 24 hour period, yield a maximum-brightness map, a minimum-brightness map, and an average-brightness map of the AR.  Running sets of such three maps come from repeating this process for each time step of running windows of 20, 16, 12, 8, 5, 3, 1 and 0.5 hours.  From each running window's set of three maps, we obtain the AR's three corresponding luminosity light curves.  We find: (1) The time-averaged ratio of minimum-brightness-map luminosity to average-brightness-map luminosity increases as the time window decreases, and the time-averaged ratio of maximum-brightness-map luminosity to average-brightness-map luminosity decreases as the window decreases.  (2) For the 24-hour window, the minimum-brightness map's luminosity is 5\% of the average-brightness map's luminosity, indicating that at most 5\% of the AR's hot94 luminosity is from heating that is steady for 24 hours.  (3)  This upper limit on the fraction of the hot94 luminosity from steady heating increases to 33\% for the 30-minute running window.  This requires that the heating of the 4--8 MK plasma in this AR is mostly in bursts lasting less than 30 minutes: at most a third of the heating is steady for 30 minutes. 

	\end{abstract}
	

	\section{Introduction} \label{sec:intro}

The temperature of solar active region (AR) corona is millions of kelvins, hundreds of times that of the photosphere, presenting a major mystery of solar-stellar astrophysics \citep{klim06,real14,hino19}. The corona in ARs is much hotter than that of the quiet Sun \citep{zirk93}, evidently due to their stronger magnetic field. Several scaling laws have been derived in which the rate of heating is directly proportional to some power of the magnetic field strength in coronal loops and/or at their photospheric feet \citep{rosn78,golu83,klim95,mand00,schr04,warr06,martens10,vanb11}. 
However, the heating in coronal loops does not depend on the magnetic field alone; convection plays an equally important role \citep{tiw17} -- so that the loops connecting the strongest magnetic field regions in ARs (umbra to umbra) are the dimmest coronal loops, challenging the general validity of scaling laws. 

Parker proposed that ``nanoflares" release magnetic energy of the order of 10$^{24}$ erg per nanoflare and that these are ubiquitous in closed-loop systems such as ARs \citep{parker88}. Nanoflares are a result of `local' energy release by magnetic reconnection.  The magnetic reconnection gets induced by braiding of magnetic field lines driven by photospheric magnetic convection \citep{parker72,parker83a}. Such reconnections are  also possible when oppositely twisted coronal loops interact \citep{gold60}.
Other heating mechanisms include heating by MHD waves generated by photospheric convection acting on the feet of coronal magnetic field, most commonly invoked for open fields \citep[e.g.,][and references therein]{oste61,vanb11}. We focus on the closed-loop AR coronal heating in this work. In ARs, again, the coronal heating can be by dissipation of MHD waves generated by magnetic reconnection. 

AR coronal heating can be high-frequency (in that the resulting coronal brightness of the heated loop is nearly steady), or, low-frequency in that cooling of the heated loop happens between each heating episode, which is ``impulsive" \citep[e.g.,][]{klim15}. There have been observational reports of both low-frequency and high-frequency heating and it has been found that the low-frequency heating is significant \citep[e.g.,][]{ugar12}.  
Magnetic reconnection as such is intrinsically impulsive or unsteady \citep{prie07}; see e.g., \cite{nitt00}, and \cite{real14} for observational support of this idea. 
Modelling results of AR coronal heating have largely suggested that the properties of observed nanoflares are arguably incompatible with nanoflare models that require long energy build-up ($>$10 s -- several 1000 s) and with steady heating \citep{carg15}.

\begin{figure*}
	\centering
	\includegraphics[trim=6cm 4cm 5.5cm 0.4cm,clip,width=0.91\linewidth]{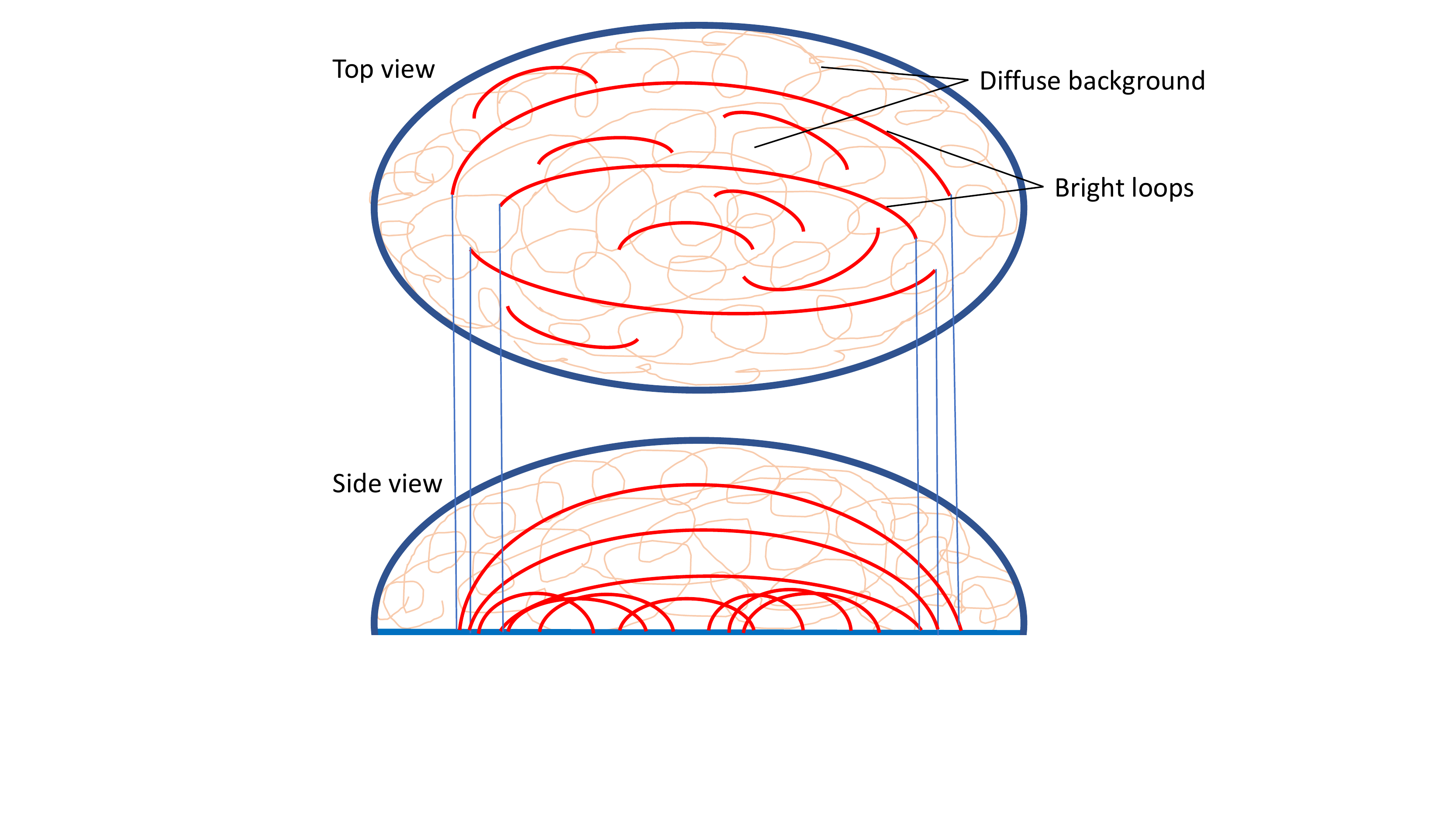}
	\caption{A drawing depicting the diffuse background and bright coronal loops in an active region (AR). The upper panel is a top view, and the bottom panel is the corresponding side view. The bright loops are in red, and the diffuse background is in light orange colour, which represents random unresolved structures. Six vertical lines connecting the top and side views, in blue, mark the ends of the three largest bright loops, for clarity. } 
	\label{fig1_cartoon}
\end{figure*}

Background heating at large scales on the Sun is often referred to as non-AR, quiet Sun, heating \citep{klim15,saku17}. 
The solar AR corona consists of multiple unresolved structures (diffuse background) and resolved structures (bright coronal loops), see Figure \ref{fig1_cartoon} for a simple illustration. The unresolved AR corona might contribute substantially, if not dominantly, to an AR's coronal radiation \citep{cirt05,asga21}. The ambient (or background, `diffuse') corona in ARs is at 1--2 MK, which is also the temperature of many bright AR loops \citep{warr02,cirt06,broo19}. It is therefore a challenge to single out and measure transient coronal loops without contamination from the diffuse background along the line of sight (LOS). We cannot understand the heating mechanism/s, observationally or in models, if we cannot determine the source of the emission, i.e., cannot determine how much of the emission comes from bright coronal loops, and how much from the diffuse background that remains unresolved in high-resolution EUV  observations \citep[e.g.,][]{cirt05}.   

Different methods have been employed in the past to figure out and correct for this LOS contamination.  \cite{test02} measured intensity profiles across loop structures and assumed the background to be at the level of the line that fits the lowest intensity values in the profile. The peaks of profiles minus the background were considered to be the intensity of coronal loops. \cite{port95} and \cite{wine03} used linear interpolation of the data pixels just outside the loop pixels and smoothed this image and then subtracted it from the original image. 
\cite{delz03} used a similar method in that they calculated intensity of emission of lines within and just outside coronal loops and subtracted the outside background intensity from the intensity within the loops. \cite{broo19} performed emission measure analysis of an AR to infer information on the diffuse (unresolved) background, finding that about half of the regions in their sample have narrow EM distributions, peaking at T = 1.4 -- 2 MK.

Here we present a different approach to quantitatively assessing background and transient coronal heating in ARs. From a set of AR coronal EUV  images -- from stepped times over a selected time interval (time window) -- we extract a minimum-brightness map, a maximum-brightness map, and an average-brightness map. Any loop-like feature in the minimum brightness map is a ``steady loop" during the time window: each pixel of this loop is steadily at least that bright in the coronal EUV image at each step time in the window. On the other hand, the brightness of each pixel of any loop-like feature seen in the maximum-brightness map is expected to come from the peak of a transient brightening during the time window. The brightness of each pixel of the average-brightness map is the time-average brightness in that pixel during the time window.

\section{Data and Methods}\label{sec:data}
	
\begin{figure*}
	\centering
	\includegraphics[trim=2.65cm 2.6cm 4.2cm 1.8cm,clip,width=\linewidth]{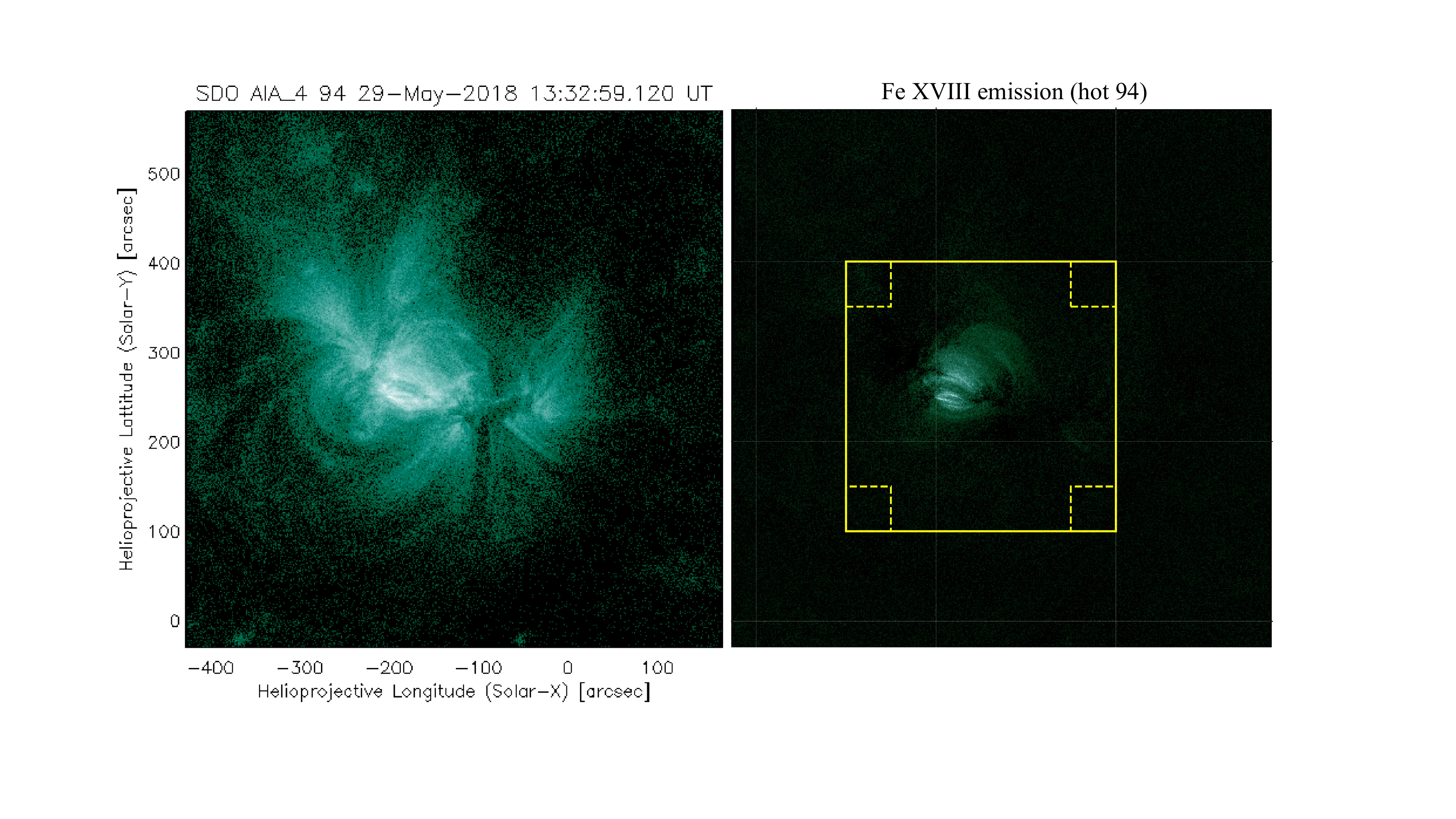}
	\caption{Snapshot of the 4--8 MK quiescent core of NOAA AR 12712. The left panel is an example image of the AR in the SDO/AIA 94 \AA\ channel. The right panel is the 4--8 MK \FeXVIII\ emission (hot component of the AIA 94 \AA\ channel) of the image in the left panel. The yellow solid box outlines the region we analyze in this research work. The four dashed boxes in four corners outline the region used for the noise estimation in hot 94 images.} 
	\label{f2}
\end{figure*}

We use 24 hours of extreme ultraviolet (EUV) images of NOAA AR 12712 observed on May 29, 2018 with the Atmospheric Imaging Assembly \citep[AIA;][]{leme12} on-board Solar Dynamics Observatory \citep[SDO;][]{pesn12}. Although AIA provides EUV images at 12-second cadence we use 3-minute cadence data, which is sufficient for our purpose in the present study. We followed the method of \cite{warr12} and calculated ``hot 94" images, i.e., \FeXVIII\ emission (radiated from the coronal plasma at 4-8 MK) by removing the warm components (of about 1 MK) of the AIA 94 \AA\ channel. For this purpose we combined AIA 94 \AA\ channel data with the data of AIA 171 and 193 \AA\ channels. This is the same data set (\FeXVIII\ emission) as used in \cite{tiw21} for finding out the mix of positive and negative magnetic flux at the feet of the brightest transient coronal loops in this AR; see their Figure 1. This AR is also the same AR as observed by Hi-C 2.1 \citep{rach19,tiw19,pane19}. In Figure \ref{f2} we display an example AIA 94 \AA\ image of the AR, as well as the extracted hot 94 (\FeXVIII\ emission) image.  

Because a significant amount of high-temperature emission in ARs can come from a decaying flare lasting many hours (extended post-flare phase can last $\ge$10 hours, e.g., \citealt{asch01}), we selected this AR that did not produce any significant flare ($\ge$ B-class GOES flares) from a day before through a day after our 24-hour observation period (the 24 hours of 2018 May 29). 
It is also worth mentioning that the AR under investigation was at the peak of its emergence during the
24-hour observations, and it started decaying soon after, or even during, our 24-hour observation period \citep{tiw21}.

We estimated the noise in the hot 94 images by calculating the 1$\sigma$ variation of the intensity inside the four boxes in the corners outlined by the yellow boxes, shown in the hot 94 image in Figure \ref{f2}, and averaging those four values. We take this averaged 1$\sigma$ intensity noise value as the lower threshold for above-noise pixels, i.e., only pixels having values above this noise level were used for our analysis.    
Thus, remaining noise, if any, should be quite small.

\begin{figure*}
	\centering
	\includegraphics[trim=6.4cm 1.2cm 7.6cm 0.5cm,clip,width=\linewidth]{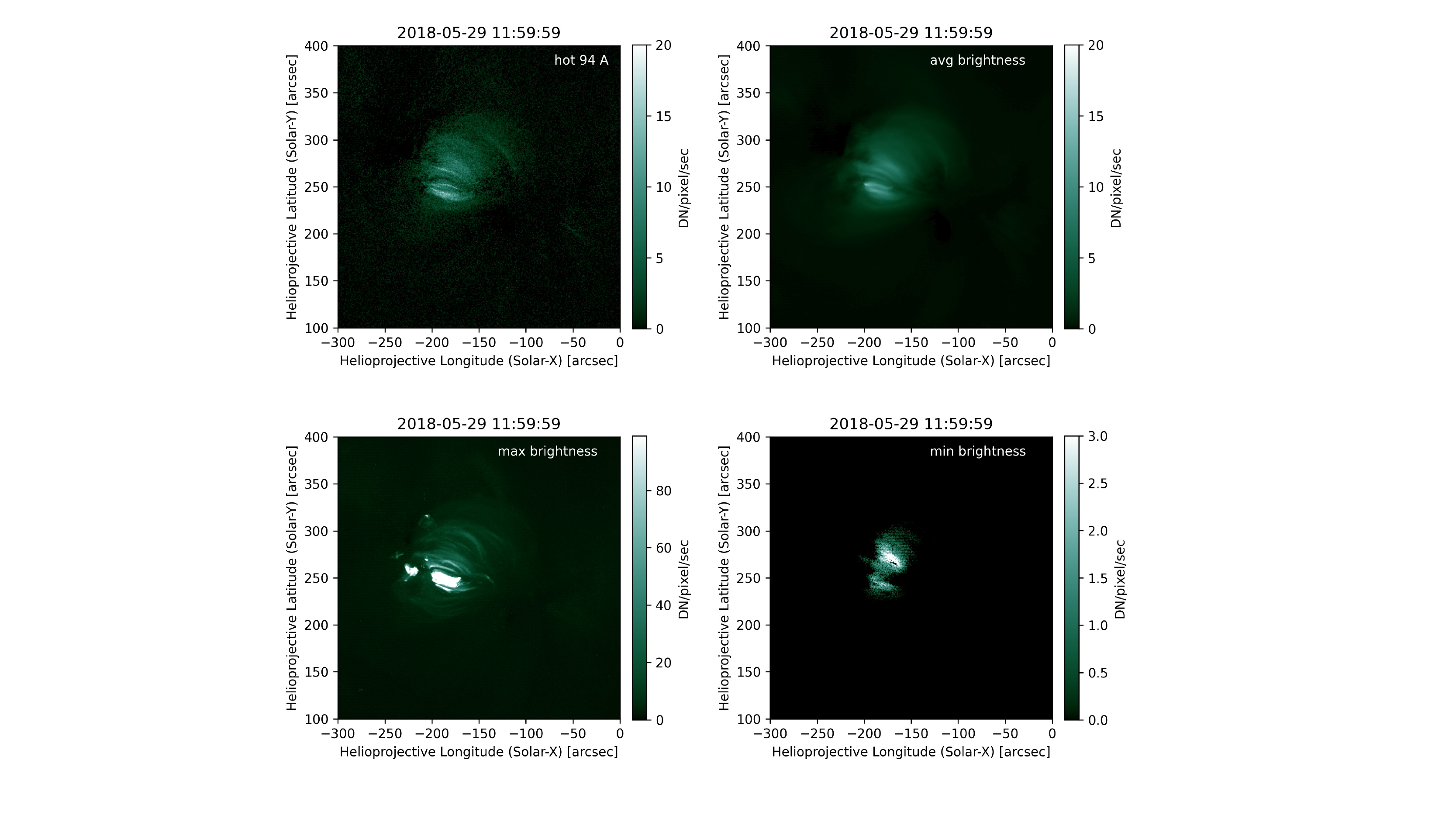}
	\caption{The four types of images discussed in the text. Starting clockwise from the top left -- an example \FeXVIII\ image frame from the 24-hour data set, the map with average pixel values over the 24 hours, the map of pixel maximum values over the 24 hours, and the map of pixel minimum values over the 24 hours. The time (11:59:59) is the median time of the images in the 24 hours.  Each image is plotted on its own intensity scaling, as noted on the color bars. Similar maps were made for each 3-minute step of the 20, 16, 12, 8, 5, 3, 1 and 0.5 hours running windows.} 
	\label{f3}
\end{figure*}

\begin{figure*}
	\centering
	\includegraphics[trim=6.5cm 1.05cm 7.5cm 0.5cm,clip,width=\linewidth]{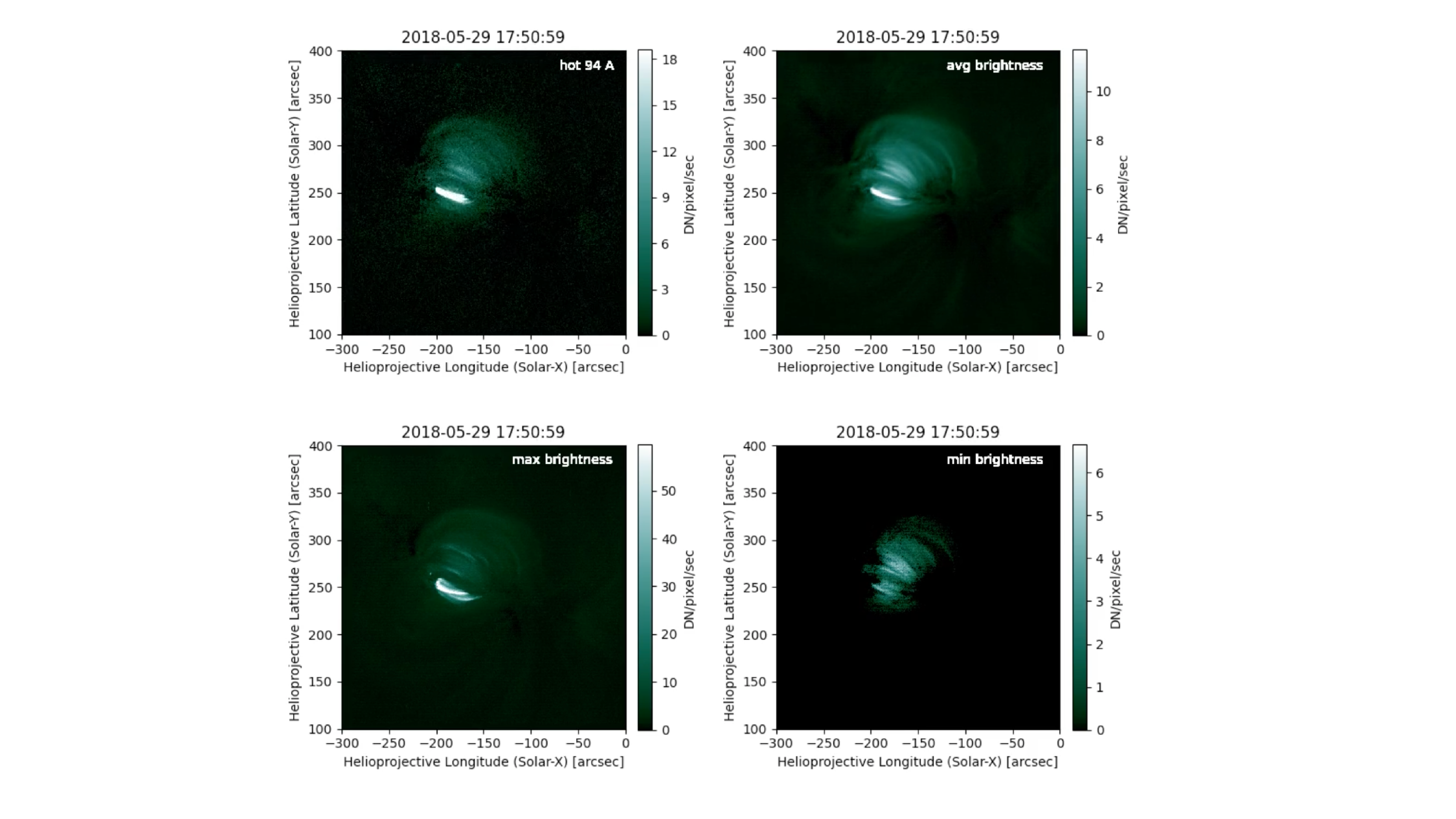}
	\caption{Same as the Figure \ref{f3}, but taken from the 3 hour running window. Starting clockwise from the top left -- an example \FeXVIII\ image frame from the 24-hour data set, an example map of pixel average values from the 3-hour running window centered at the same time (17:50:59 UT), an example map of pixel maximum values from the 3-hour running window centered at the same time, and an example map of pixel minimum values from the 3-hour running window centered at the same time. The time (17:50:59) is the central time for the 3-hour window to which these images belong. As evident from the color bars, each image has its own intensity scaling; see the movie ``3hr\_4panel.mp4". Similar maps were made for the 24, 20, 16, 12, 8, 5, 1 and 0.5 hours running windows. } 
	\label{f4}
\end{figure*}

\begin{figure*}
	\centering
	\includegraphics[trim=3cm 1.1cm 17cm 0cm,clip,width=0.88\linewidth]{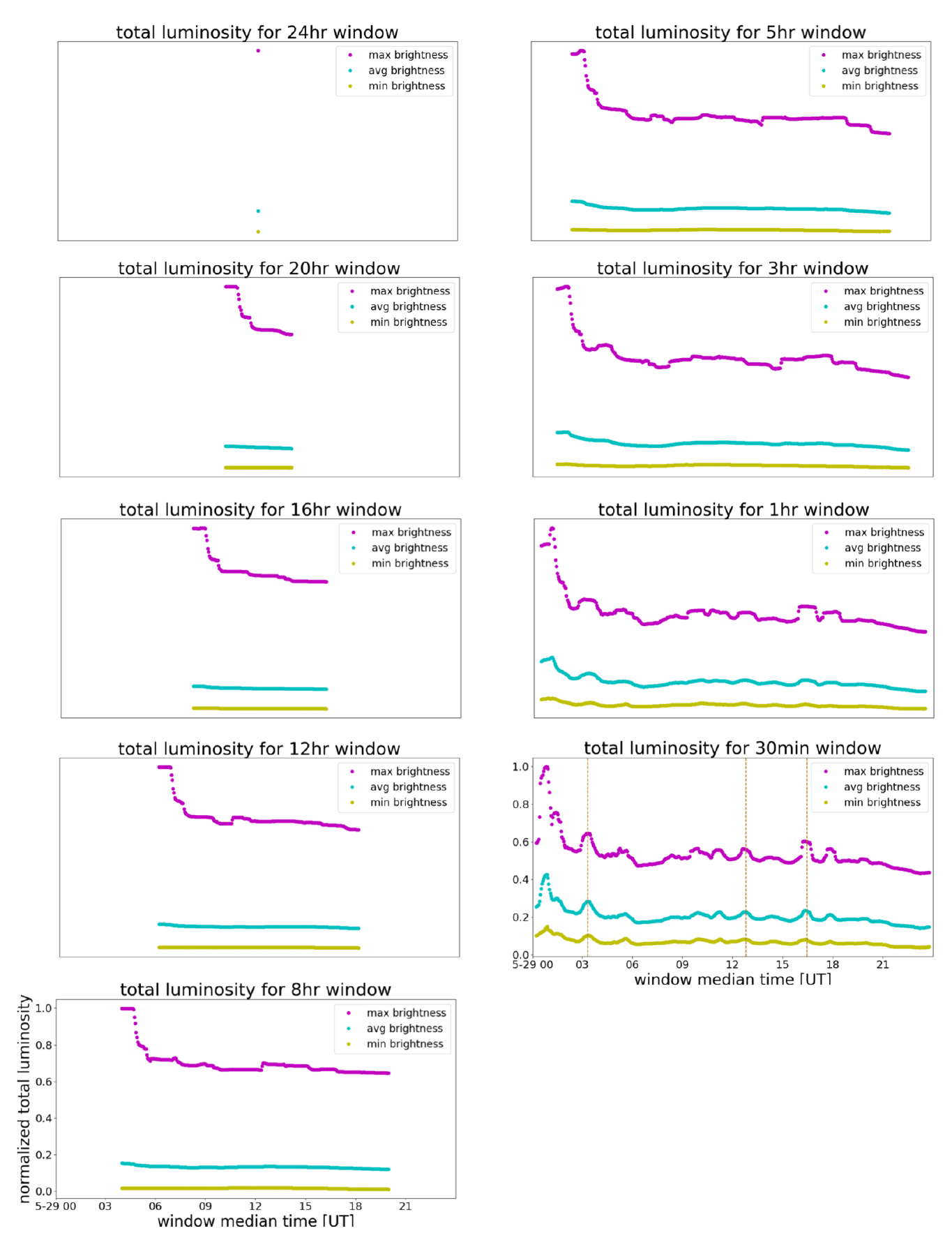}
	\caption{Normalized light curves of minimum (L$_{min}$), maximum ( L$_{max}$) , and average brightness ( L$_{avg}$) maps for all nine time windows, the windows being progressively smaller from top to bottom. The normalization is done by dividing each light curve with the maximum value of the light curve of the maximum brightness map. The magenta, cyan, and chartreuse coloured light curves are for maximum, minimum, and average brightness maps, respectively. On the X-axis, the window's median time is given in the format of ``05-29 00", ``05-29 03"...etc., which means 29-May-2018 00 UT, 29-May-2018 03 UT...etc. The time stamps are at every 3 hours.
		As obvious, for the 24-hour window each light curve has a single point. As the time interval (running time window) decreases the data points increase in number. Each data point is separated by 3 minutes from the next or earlier data point. In the 30-minute light curves plot we show correlation between the three light curves at three local intensity peaks via three dotted vertical lines.
	}
	\label{f5}
\end{figure*}

\begin{figure*}
	\centering
	\includegraphics[trim=3cm 1cm 17cm 0cm,clip,width=0.91\linewidth]{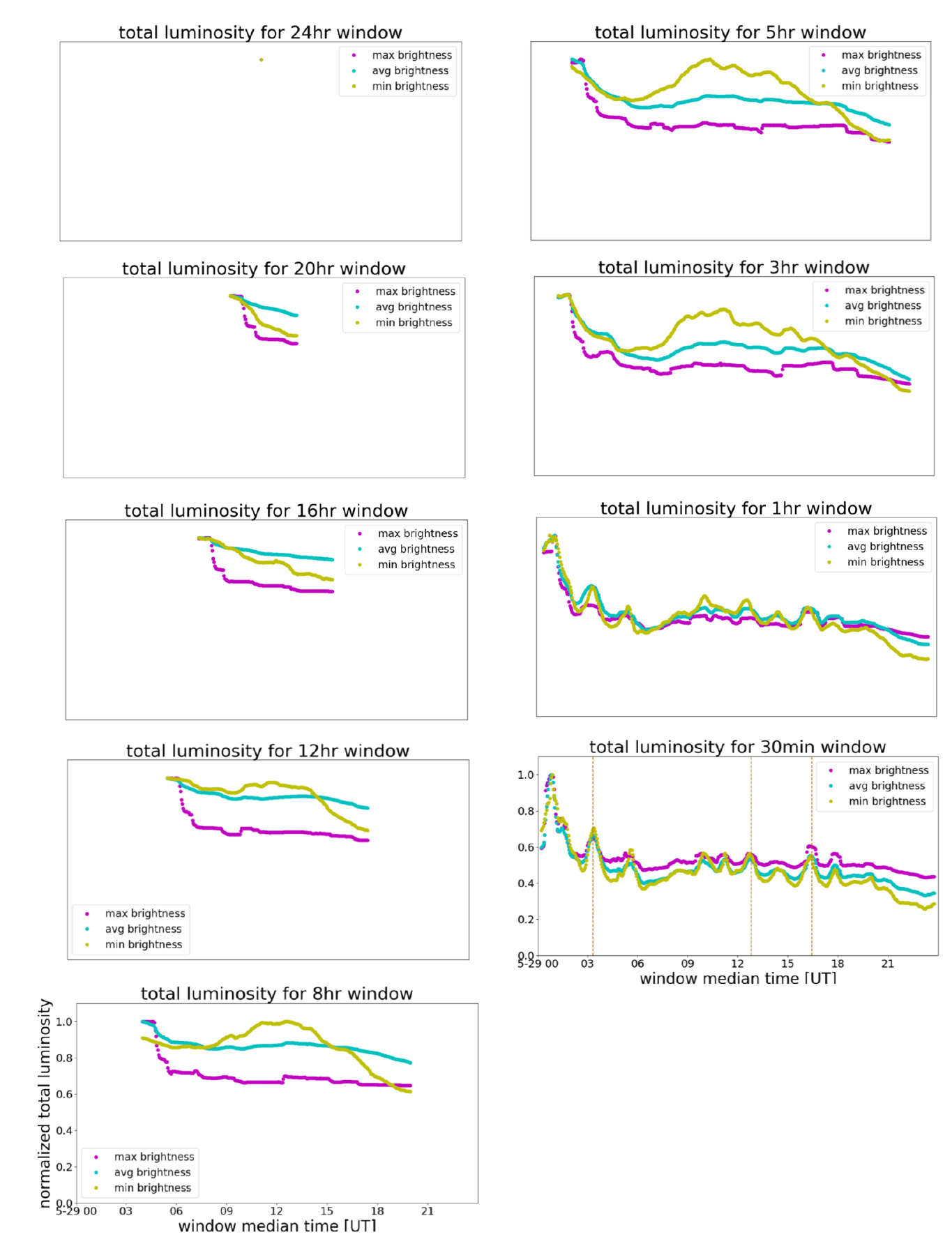}
	\caption{Same as Figure \ref{f5}, but here normalization is done by dividing each light curve with its own maximum value.
	}
	\label{f6}
\end{figure*}

\begin{figure*}
	\centering
	\includegraphics[trim=2.6cm 0.55cm 16cm 0cm,clip,width=0.89\linewidth]{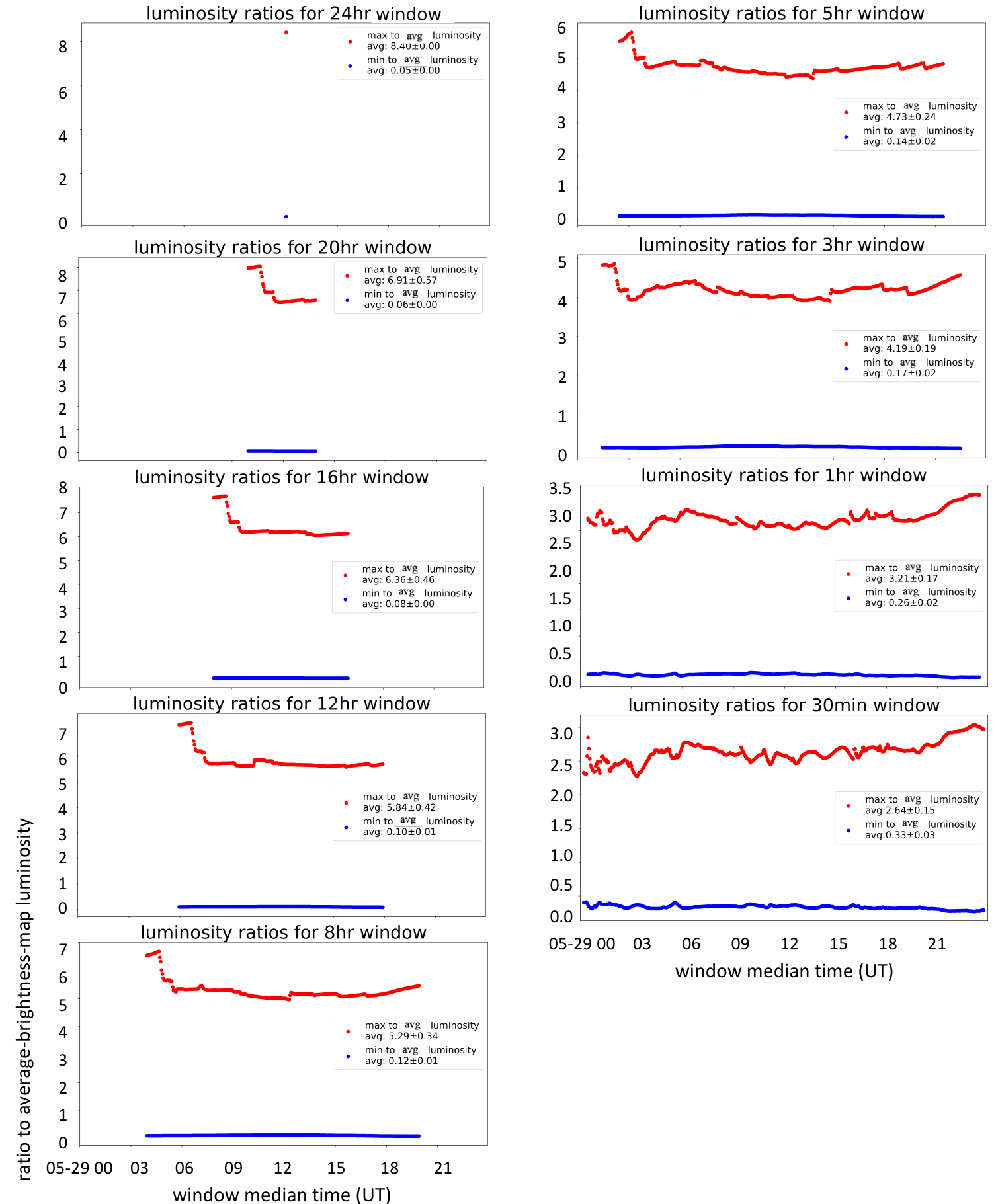}
	\caption{Luminosity ratio plots for all the nine time windows progressing from the 24-hour window to the 30-minute window from top to bottom. Red and blue colours are for the ratio of the maximum-brightness-map luminosity to the average-brightness-map luminosity (R$_{max/avg}$) and the ratio of the minimum-brightness-map luminosity to average-brightness-map luminosity (R$_{min/avg}$), respectively. The X-axis is the window's median time, as in Figures \ref{f5} \& \ref{f6}. 
	} 
	\label{f7}
\end{figure*}

To quantify the steady background heating we create minimum brightness maps in the following way. We find the minimum brightness of each pixel in the hot 94 images in nine different time windows, which range from the 24 hours to 30 minutes in width. For the 24 hour window a single map is obtained containing the minimum intensity at each pixel of the map over the 24-hour full sequence of 3-minute-cadence 480 hot 94 images. For the 30-minute window we find the minimum intensity of each pixel in the first 30-minute window (centered on the fifth image from the start of the 480 image sequence) and then in the next 30-minute window (centered on the sixth image of the 480 image sequence), and so on. This provides a series of running minimum brightness maps at a 3-minute cadence. In the same way we obtained running minimum brightness maps for different time windows: in total for nine windows of 24, 20, 16, 12, 8, 5, 3, 1, and 0.5 hours. The minimum-brightness map for the full 24-hour interval can be seen in Figure \ref{f3}.  A similar method was used by \cite{falc97} to find ``persistently bright" coronal loops i.e., the long lasting steady bright loops in ARs. 

Further, to quantify the bursty, transient heating, we made maximum-brightness maps for all those intervals used for creating minimum-brightness maps. For each running time window, for making maximum-brightness maps, instead of finding minimum intensity at each pixel over the time of the span of the window for each step of the window, we found the maximum intensity at each pixel. The maximum-brightness map for the 24-hour interval can be seen in Figure \ref{f3}. Note that, as mentioned before, the selected AR is a quiescent one with no large flares. Therefore, we do not expect, neither did we notice, any of the image pixels to be saturated.   

In addition to making minimum-brightness maps and maximum-brightness maps of \FeXVIII\ emission of the AR, we also made average-brightness maps for all the nine time intervals. The average-brightness map contains the mean value of each pixel over the time window.
The average-brightness map luminosity is used in interpreting the luminosities of the minimum- and maximum-brightness maps, described in the next paragraph. Thus, in the same way as for minimum-brightness and maximum-brightness maps, there is only one average brightness map for the 24-hour window, and there are progressively more average-brightness maps for progressively shorter running time windows.

A map's luminosity is the pixel-sum total brightness of the map. For each time step of each time window, we have the three maps (the minimum-brightness map, the maximum-brightness map, and the average-brightness map), the three luminosities (the minimum-brightness-map luminosity L$_{min}$, the maximum-brightness map luminosity L$_{max}$, and the average-brightness-map luminosity L$_{avg}$), and their ratios (R$_{min/avg} = L_{min}/L_{avg}, R_{max/avg} = L_{max}/L_{avg},$ and $R_{max/min} = L_{max}/L_{min}$).  For each step of the time window, the brightness of each pixel of the each of the AR's ``hot 94" images in the time window is the sum of two components: (1) l$_{steady}$, the emission from all of the resolved and all of the unresolved structures along the line of sight through that pixel for which the ``hot 94" emission along that line of sight is constant during the time interval (of length $\tau_w)$ spanned by the window, and (2) l$_{bursty}$, the emission from all of the resolved and all of the unresolved structures along the line of sight for which the ``hot 94" emission along that line of sight is not constant (is bursty) during the time interval spanned by the window.  Therefore, at each time step of the time window, L$_{min} \tau_w$ is an upper bound on the AR's ``hot 94" emission that is steady during the time spanned by the time window, L$_{avg} \tau_w$ is the AR's total ``hot 94" emission (steady plus bursty) during the time interval spanned by the time window, and R$_{min/avg}$ is an upper bound on the fraction of the AR's total ``hot 94" emission that is from the steady component of that emission during the time interval spanned by the time window.  If only some of the AR's ``hot 94" emission is constant during the time interval spanned by the time window, then, because part of $L_{min} \tau_w$ is from the AR's bursty ``hot 94" emission, 1 $- R_{min/avg}$ is a lower bound on the fraction of the AR's total ``hot 94" emission that is from the bursty component of that emission during the time interval spanned by the time window.

We obtain the three map luminosities for each step of each time window to obtain the run of luminosity of each of the three kinds of maps with time over the time range sampled by the running time window, i.e., to obtain luminosity light curves. From the map-luminosity light curves from each running time window, we obtain (i) the time-average of R$_{max/avg}$ (the time average of the ratio of the luminosity of the maximum-brightness map to the luminosity of the average-brightness map), and (ii) the time average of R$_{min/avg}$ (the time average of the ratio of the luminosity of the minimum-brightness map to the luminosity of the average-brightness map).  From that, we obtain (i) the dependence of these two time-averaged ratios on the width of the running time window, and (ii) for each running-window width, (1) a time-averaged upper bound on the fraction of the total heating of the AR's \FeXVIII-emission plasma that could be steady instead of bursty on the time scale of the window width (i.e., time-averaged $R_{min/avg}$), and (2) a time-averaged lower bound on the fraction of the total heating of the AR's \FeXVIII-emission plasma that could be bursty instead of steady on the time scale of the window width (i.e., 1$-$time-averaged $R_{min/avg}$).  

Figure \ref{f3} shows a four panel image in which an example hot 94 image and the average-brightness, maximum-brightness, and minimum-brightness maps for the 24-hour window are displayed. In Figure \ref{f4} we show another similar example but this time from a running time window for the 3-hour time interval. 
A movie corresponding to Figure \ref{f4} shows the evolution of these maps, and is available online.

\section{Results} 
	
In Figure \ref{f5} we display normalized light curves of the three types of examined maps i.e., minimum-brightness, maximum-brightness, and average-brightness maps, for nine different running time windows. As is obvious, the 24-hour window light curves have a single point for each of the three cases. 
In general, the fluctuations in each of the three light curves increase as the time window decreases. This is reasonable because the narrower the running time window, the more the short-time heating activities are isolated.
However, we have not calculated the lifetime of individual events. 
Therefore, how the map brightness for each time-window changes with the event lifetime is not addressed in the present work.

The Figure \ref{f5} shows luminosities for the three cases relative to each other, because the normalization is done by dividing each light curve with the maximum value of the light curve of maximum brightness maps. To find out true fluctuations in each of the three light curves, we normalised each of the three light curves individually to their own maximum values, and show in Figure \ref{f6}.

The beginning of particularly the maximum-brightness-map light curves, for about three hours, is notably luminous, due to enhanced heating activities in the beginning of our data set, when the AR was more newly emerged. The luminosity in general decreases over time. This is consistent with the progression of the AR's coronal heating seen in the Movie1 (3hr\_4panel.mp4): the AR becomes less active with time after completion of emergence. 

Based on Figure  \ref{f5}, the minimum-brightness-map light curves apparently show the least variations over time in the different running windows. The light curves for the average-brightness maps apparently fluctuate more than the light curves for the minimum-brightness maps and fluctuate less than the light curves for the maximum-brightness maps. However, as mentioned before, these fluctuations are relative. To find out the true fluctuations in each of these light curves we normalized each plot with their own maximum values (see Figure \ref{f6}). 
Figure \ref{f6} shows that the fluctuations in all the three plots are not too different. This was not apparent in Figure \ref{f5}.  

The luminosity enhancements/transients in all the three plots strongly correlate in time. This correlation becomes more evident with decreasing time window width, and becomes most obvious in the 30-minute window. This correlation is qualitatively demonstrated in Figure \ref{f5} by drawing three vertical dotted lines in the luminosity plots for the 30-minute window. This indicates that when the maximum-brightness map luminosity changes the minimum-brightness map luminosity changes as well. In other words, the ``background" intensity also gets enhanced with a transient event. This would be the case if most of the ``background" heating were simply the decay tails of transient heating events. 

To quantify the relative strength of background and transient heating, in Figure \ref{f7}, we display the plots of the ratio of the luminosity of the maximum-brightness map to the luminosity of the average-brightness map and the ratio of the luminosity of the minimum-brightness map to the luminosity of the average-brightness map, for all the nine time windows. These plots present the temporal variation of these two map-luminosity ratios. Again, the greater scatter in the points in the beginning of the plot for the 30-minute running window for the ratio of the maximum-brightness-map luminosity to the average-brightness-map luminosity shows briefer brightenings, i.e., resolved shorter-lifetime transients, than do the beginnings of the corresponding plots for the wider windows.

The time-averaged ratio of the luminosity of the maximum-brightness map to the luminosity of the average-brightness map comes out to be 8.40$\pm$0.00, 6.91$\pm$0.57, 6.36$\pm$0.46, 5.84$\pm$0.42, 5.29$\pm$0.34, 4.73$\pm$0.24, 4.19$\pm$0.19, 3.21$\pm$0.17, and 2.64$\pm$0.15, for 24, 20, 16, 12, 8, 5, 3, 1, and 0.5 hours windows, respectively. That is, the \FeXVIII\ maximum-brightness-map luminosity from transient heating is several hundred percent more than the \FeXVIII\ average-brightness-map luminosity for each window, the maximum being $\sim$850\% for the 24-hour window and the minimum being $\sim$260\% for the 30-minute window. 

The time-averaged ratio of the luminosity of the minimum-brightness map to the luminosity of the average-brightness map comes out to be 0.053$\pm$0.00, 0.06$\pm$0.00, 0.08$\pm$0.00, 0.10$\pm$0.01, 0.12$\pm$0.01, 0.14$\pm$0.02, 0.17$\pm$0.02, 0.26$\pm$0.02, and 0.33$\pm$0.03 for 24, 20, 16, 12, 8, 5, 3, 1, and 0.5 hours windows, respectively. This indicates that, for the AR's \FeXVIII-emission plasma, the ratio of the background (steady) heating rate to the average heating rate is small, at most $\sim$5\% for the 24-hour window, and at most $\sim$33\% for the 30-minute window.  

\begin{figure*}
	\centering
	\includegraphics[trim=2.5cm 1cm 2.5cm 2.5cm,clip,width=\linewidth]{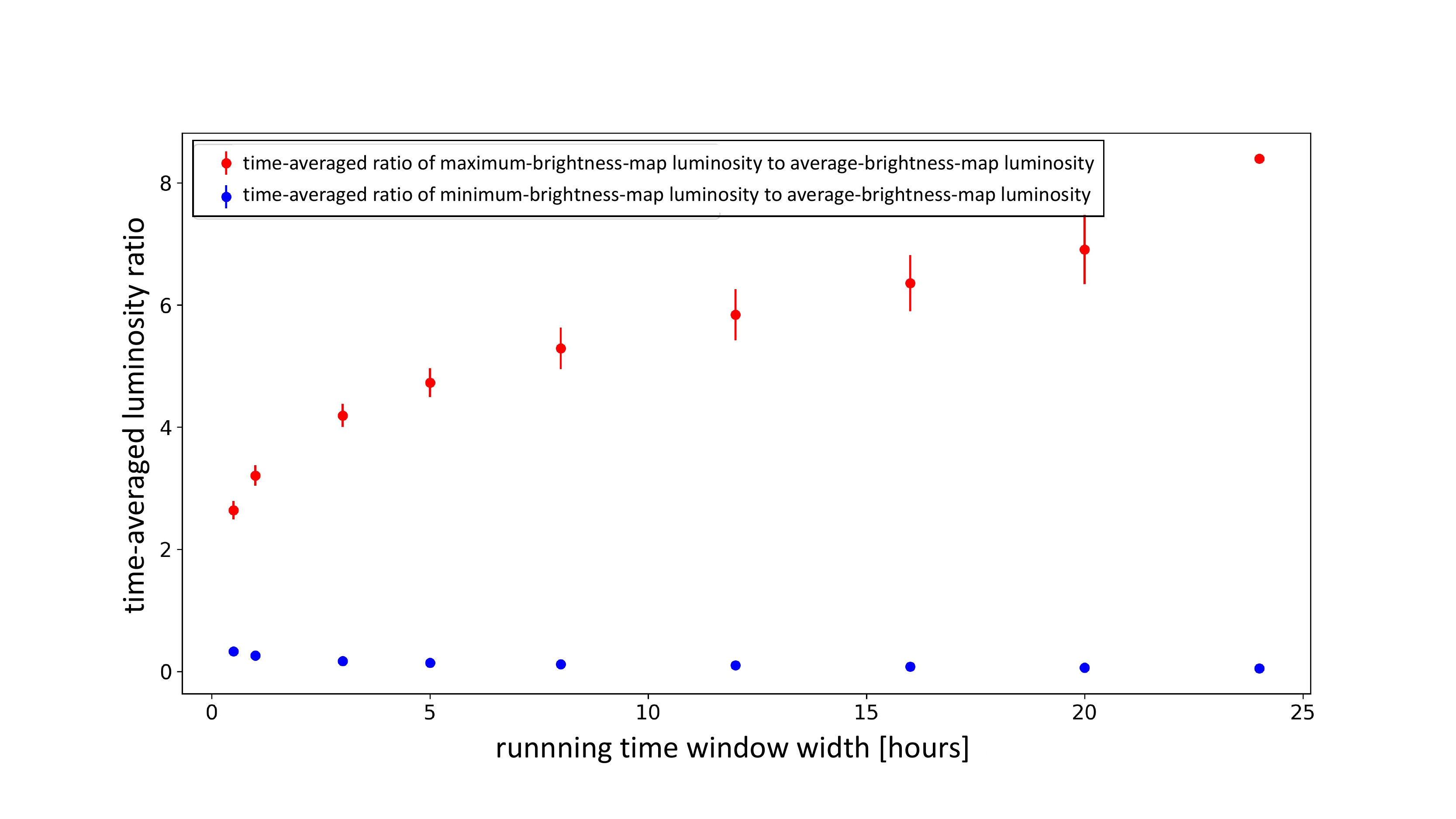}
	\caption{Plots displaying how the time-averaged ratios of minimum-brightness-map luminosity to average-brightness-map luminosity (in red) and maximum-brightness-map luminosity to average-brightness-map luminosity (in blue) depend on the width of the running time window. As expected, the time-averaged ratio of minimum-brightness-map luminosity to average-brightness-map luminosity (time-averaged R$_{min/avg}$, in blue) increases as the time window decreases, and the time-averaged ratio of maximum-brightness-map luminosity to average-brightness-map luminosity (time-averaged R$_{max/avg}$, in red) decreases as the time window decreases. The standard deviations for blue are barely visible because they are so small.  } 
	\label{f8}
\end{figure*}

Further, to graphically display the trends of the above ratios with time-window width, Figure \ref{f8} plots for each time-window width the time-average ratio of the maximum-brightness map's luminosity to the average-brightness map's luminosity and the time-average ratio of the minimum-brightness map's luminosity to the average-brightness map's luminosity. The numbers used in these plots are also printed on each panel of Figure \ref{f7}  for each time window, and are the time-averaged values of the luminosity plots in that figure. Thus, to reiterate, the values in the plots of Figure \ref{f8} quantify the relative strengths of the background and transient heating.

As the time-window width increases, the time-average ratio of maximum-brightness-map luminosity to average-brightness-map luminosity obviously increases.      
The ratio of minimum-brightness-map luminosity to average-brightness-map luminosity decreases with increasing time-window width but at a much smaller pace than the increase in the ratio of maximum-brightness-map luminosity to average-brightness-map luminosity with increasing time-window width. It particularly does not show a significant change for increasing time-window width beyond 5 hour. This suggests that there might be background or steady heating of the plasma, although as only a small fraction of the total heating.

\section{Summary and Discussion}
	
We investigate the background and transient heating of the \FeXVIII-emission hot coronal plasma of an active region (AR) and quantitatively assess these using running time windows ranging in width.
The use of minimum-brightness maps, maximum-brightness maps, and average-brightness maps is a new technique to gauge the background coronal heating and transient coronal heating in ARs (in the way presented here for 4--8 MK coronal plasma of one AR).
The 24 hour-window results indicate that at most 5\% of the \FeXVIII\ luminosity of the AR on average comes from heating that is steady for 24 hours. This upper limit increases to 33\% for the 30-minute running window.

We  use \FeXVIII\ emission because we are interested in quantifying the background heating of the hottest emission in a narrow wavelength, avoiding noise or false heating via contamination of other wavelengths, as usually happens for an AIA channel. In particular, it is well known that AIA 94 \AA\ channel contains emission from the plasma at 1 MK and 4--8 MK \citep{leme12,warr12}. Therefore we isolate only the hot emission from AIA 94 \AA\ channel and apply our method to quantify the transient and background heating of the AR's \FeXVIII-emission coronal plasma at 4--8 MK.

We note that in the present work we quantify the background and transient heating via time-averaged ratios. The purpose here is not to quantify energies in individual events. Rather it is to find out, on average, what percentage of the average AR heating might come from steady background, and what percentage of the average AR heating might come from transient activities.

Effectively, our assumption in this method of quantifying the background  (steady) heating is that the AR's unresolved diffuse-background hot-94 corona does not vary significantly during the interval of the running time window at each step of the window, and that during that interval the brightness of each resolved bright coronal loop does significantly evolve.  Under this assumption, the minimum-brightness map captures an upper-bound approximation of the AR's ``steady" background hot corona for time window's interval at each step of the window.  More exactly, we assume that the AR's hot corona has a steady component that is steadily sustained by corresponding steady heating.  Under this assumption, the brightness of each pixel of each minimum-brightness map is an upper bound on the brightness from steady heating in that pixel during that map's time interval.  This interpretation of the minimum-brightness map is true even if there is zero steady heating and no steady component of the AR's hot corona.

We selected the time window of 24 hours as the upper limit in our study because to best of our knowledge no transient coronal loops reported in the literature, or in our selected AR, live more than 24 hours (so that the background steady emission could dominate in minimum brightness maps). Most AR coronal loops typically last $\sim$20–-40 minutes \citep{mulu11,pete12,tiw21}. Nonetheless, AR coronal loops have been observed to live as short as a few minutes \citep{wine13,tiw19} to as long as several hours \citep{lope2007,klim10}. The loops that last several hours are most likely due to mutual interactions of several to many loops and thus are due to repeated adjacent heating episodes  \citep{warr02,tiw14}. The hot loops in our AR have on average a lifetime of about 45 minutes \citep{tiw21}. Therefore, selecting a window width of less than 30 minutes would be unreasonable because that would include more bright loops in the background. 

We used running time windows ranging in width from 24 hours (the full span of our \FeXVIII\ image sequence) through eight progressively narrower windows, down to 30 minutes. 
Each window, via the minimum-brightness and average-brightness maps, yields an upper limit on the steady component of the AR's \FeXVIII\ emission and a lower limit on the bursty component of that emission.  That is, for any given running time window, at each time step of the window, the ratio R$_{min/avg}$ of the minimum-brightness-map luminosity to the average-brightness-map luminosity is an upper bound on the fraction of the AR's \FeXVIII-emission luminosity that is steady during the time interval of the window, and 1 $-$ R$_{min/avg}$ is a lower bound on the fraction of the AR's \FeXVIII-emission luminosity that is not steady (i.e., is bursty) during the time interval of the window.
As mentioned above, coronal loops in ARs are typically transient, living a few minutes to a few hours \citep[e.g.,][and references therein]{tiw21}. Here we quantify the quasi steady background heating and transient heating of the 4--8 MK plasma in a solar AR by following it in \FeXVIII\ emission for 24 hours.     
	
The background steady heating is expected in the \cite{parker88} scenario by magnetic energy dissipation via magnetic reconnection of fine-scale braided magnetic flux tubes within and between an AR's resolvable coronal loops. These smaller events, which occur
much more often, each contain significantly less energy than a loop-brightening event, and presumably form the diffuse background heating. 

The wave-heating coronal heating model of \cite{vanb11} is also for coronal heating that is predominantly steady, which is not what our observational results imply for the \FeXVIII-emission plasma in our AR. It is still quite possible that such steady-heating scenarios are valid for our AR's \FeXVIII-coronal steady heating, which we find to be small, at most a third of the total heating. Our results suggest that most of the heating energy is stored in the magnetic field in coronal loops that remain dim until a flare-like burst of magnetic energy release -- presumably via reconnection of the field -- is triggered somehow.

Flares that each release orders of magnitude more energy than does a $\le 10^{24}$ erg nanoflare have a 
power-law distribution of event frequency with event energy. In a log-log plot, that distribution has a slope that is flatter than $-$2. This shows that AR coronal heating is not done by flare events in the small-energy end of the distribution, because their power output is less than that from events in the large-energy end of the distribution, and the time-averaged power of the large-energy events is far less than is needed to power AR coronal heating \citep{huds91}.  Therefore, the bursty \FeXVIII-emission plasma in our AR is not from events that fit the flat power-law distribution of normal (larger) flares. The bursty heating events in our AR must be far more frequent than expected from the flat power-law distribution of flare events. That much is clear from our results, but our results do not determine the (event frequency) -versus- (event energy) distribution of AR's bursty heating events.

Our method to quantify diffuse background (or steady) heating filters signal from any transient heating in the given time window. We have applied this technique to only hot 94 maps, but in principle this will work on any image-series  of EUV channel or any other wavelength to obtain minimum-brightness maps and maximum-brightness maps. 
We limited our work to hot 94 because, as previously mentioned, we wanted to quantify the background heating for this AR's 4--8 MK \FeXVIII\ plasma, and introduce the technique. 
More work is needed to find out if similar portions of steady heating and transient heating are obtained in other AIA wavelengths.  We have only a first result from only one AR. 

There are several limitations to the current study. An important limitation to the quantification of diffuse background and transient heating by this method is that the effects of LOS are still there; they can particularly dilute the measurement of diffuse background heating. That means when we select the minimum value of a pixel, there always may be one or another bright loop along the LOS, obstructing us to evaluate the true background heating (as demonstrated in Figure \ref{fig1_cartoon}). Thus, our estimated steady heating values are upper limits on the background heating; the true background is even less. What we have found for hot 94 may not be true for other wavelengths; results may well be quite different for other AIA channels, and need to be investigated.

In conclusion, our results suggest that most of AR high-temperature heating is transient -- the background or steady heating is small, giving no more than a third of the total \FeXVIII\ luminosity of our AR at a given time. 
The AR's radiation output at 1--2 MK could be a lot bigger than the hot 94 emission and is not addressed here.
Whether the heating of the 1--2 MK plasma in ARs is significantly more steady and less bursty than the heating of the 4--8 MK plasma remains an open question. The presented method of creating minimum-brightness, maximum-brightness, and average-brightness maps can be utilized for other science analysis such as for finding the coolest and hottest structures, and how  these structures evolve, in flaring arcades/structures and in filament eruptions etc.

We appreciate positive remarks and constructive comments from the referee on our work.
The work reported in this paper was sparked partly by a discussion that S.K.T. and R.L.M. had with Hugh Hudson after a lecture by S.K.T. on the magneto-convection origins of AR coronal heating given at the SDO Science Workshop in 2021.
S.K.T. and R.L.M. gratefully acknowledge support by NASA HGI award 80NSSC21K0520.
L.A.W. was supported by funding from NSF grant AGS-1950831 for the UAH-CSPAR/NASA-MSFC Heliophysics REU program.
N.K.P's research was supported by NASA grants NNG04EA00C (SDO/AIA) and 80NSSC20K0720 (HGI).
AIA and HMI are instruments onboard the Solar Dynamics Observatory, a mission for NASA's Living With a Star program. The AIA and HMI data are courtesy of NASA/SDO and the AIA and HMI science teams. This research has made use of NASA's Astrophysics Data System and of IDL SolarSoft package.


\end{document}